\documentclass[prb,twocolumn,superscriptaddress,showpacs,floatfix]{revtex4-1}
\usepackage{amssymb,amsmath}
\usepackage{braket}
\usepackage{graphicx}

\begin{document}
\title{Correlated electronic structure and chemical bonding of Ce pnictides and $\gamma$-Ce} 
\author{Mikhail S. Litsarev}
\affiliation{Department of Physics and Astronomy, Uppsala University, Box 516, SE-75120, Uppsala, Sweden}
\author{Igor Di Marco}
\affiliation{Department of Physics and Astronomy, Uppsala University, Box 516, SE-75120, Uppsala, Sweden}
\author{Patrik Thunstr{\"o}m}
\affiliation{Department of Physics and Astronomy, Uppsala University, Box 516, SE-75120, Uppsala, Sweden}
\author{Olle Eriksson}
\affiliation{Department of Physics and Astronomy, Uppsala University, Box 516, SE-75120, Uppsala, Sweden}

\date{\today}

\begin{abstract}
We present calculated spectral properties and lattice parameters for
cerium pnictides (CeN, CeP, CeAs, CeSb, CeBi) and $\gamma$-Ce, within the LDA/GGA+DMFT (local density approximation/generalized gradient approximation + dynamical mean field theory) approach.
The effective impurity model arising in the DMFT is solved
by using the spin-polarized T-matrix fluctuation-exchange (SPTF) solver
for CeN compound, and the Hubbard~I (HI) solver 
for CeP, CeAs, CeSb, and CeBi. For all the addressed compounds the calculated spectral properties are in reasonable agreement with measured photoelectron spectra at high binding energies. At low binding energies the HI approximation does not manage to capture the Kondo-like peak observed for several of the Ce-pnictides. Nevertheless, the calculated lattice constants are in a good agreement with available experimental data, showing that the such a peak does not play a major role on the bonding properties. Furthermore, the HI calculations are compared to a simpler treatment of the Ce $4f$ electron as core-like in LDA/GGA for CeP, CeAs, CeSb, and CeBi, and the two approaches are found to give similar results. 


\end{abstract}

\pacs{71.20.Eh, 71.15.Nc, 71.28.+d, 71.15.Mb}

\maketitle

\section{Introduction}
Cerium monopnictides, CeX, where X $=$ (N, P, As, Sb, Bi) 
form in a rock-salt type crystal structure.
In this structure the Ce atom is located on an fcc Bravais lattice,
with the pnictogen atoms entering octahedral voids of this lattice. 
Hence, the position of the Ce atoms in the Ce pnictides is the same
as for the $\alpha$ and $\gamma$ phase of elemental cerium,
which has attracted so much attention due to its isostructural volume 
collapse~\cite{johansson,allen,koskimaki}. 
The possibility of tuning the $4f$~wave function overlap between the 
cerium atoms is an opportunity in the Ce pnictides, since the 
pnictogen atoms all have different size,
and hence they expand the fcc lattice to different degrees.
This offers a possibility to in a more detailed way investigate 
localization -- delocalization transitions in Ce based materials.

The Ce pnictides have indeed been studied intensively in the past,
due to that they display several unique and interesting properties, 
which will shortly be reviewed below. The lattice constants 
have been determined, and they suggest that the electronic structure and 
chemical bonding of CeN stand out as being different from the rest of the
series~\cite{Didchenko1963, tsymbal}. 
This has previously been suggested to be caused by itinerant
$4f$~states for CeN~\cite{patthey1, patthey2}, and the theoretical 
interpretation~\cite{delin} of the measured optical properties
are in line with this reasoning~\cite{schoenes}. 
The heavier Ce pnictides have lattice constants, and other physical properties,
which suggest a localized and non-bonding $4f$~shell.
The observed properties of the heavier Ce pnictides are in many 
ways unique, for instance, the magnetism and the crystal field splitting
of the $4f$~shell exhibit puzzling behaviour, as analyzed in detail 
in Refs.~\onlinecite{birgenau,cooper1,cooper2,rossa,vachter}.
 
Another unique feature of the Ce pnictides is found in the excited state 
properties, as revealed by the x-ray photoelectron spectroscopy (XPS). 
Several investigations have been reported \cite{xps12,xps1,xps2,xps3,xps4,xps5,xps6}, 
and they reveal an XPS spectrum which is rather 
complex as concerns the $4f$~signal. 
A feature between 2.9 - 3.2~eV binding energy is observed together
with a feature closer to the Fermi level, at around 0.6 eV binding energy. 
The exact positions of these two distinct features vary from 
pnictide to pnictide, see e.g. Ref.~\onlinecite{xps1,xps2}. 
The presence of these two peaks is at first glance surprising,
having in mind that a localized $f^1$~configuration of the heavier Ce pnictides, 
is expected to result in one spectral feature corresponding to a $f^1 \to f^0$ transition.
However, it has been pointed out~\cite{xps2,GSmodel} that two screening 
channels of the final state core hole are possible for this excitation, a well 
screened and a poor screened channel, which results in two distinct 
spectral features. The presence of two peaks in the $4f$ spectrum is a property
not only of Ce pnictides, but  also other Ce compounds (see, for example, 
Refs.~\onlinecite{twofpeak1,twofpeak2,twofpeak3,twofpeak4,twofpeak5}).

On the theoretical side, electronic structure calculations have been 
published for CeN, analyzing the energy band structure and the optical
properties~\cite{delin}. Results of this work are based on the LDA/GGA
and, essentially, correlation effects are taken into account only weakly. 
A recent calculation~\cite{cen2011Olle} for CeN focused on the electronic 
structure and the lattice dynamics. In this work the $4f$ states were considered
as delocalized, and the calculations used both the LDA/GGA level of approximation 
as well as the so called GW~\cite{gw10} approximation. 
For the heavier Ce pnictides a theory of the crystal field splitting was
proposed~\cite{wills}, which successfully reproduced the 
observations of Ref.~\onlinecite{birgenau}. The work of Ref.~\onlinecite{wills} was a combination
of model considerations and first principles theory, in which the 
localization of the $4f$ wave function was enforced in the calculation, 
and then the hybridization to other valence states was 
estimated from this electronic configuration.

Previous DMFT investigations of Ce pnictides are presented
in Refs.~\onlinecite{dmft1CePn,dmft2CePn,dmft3CePn}, wherein the Anderson 
impurity problem was solved in a finite-U extension of the non-crossing
approximation (NCA), including some lowest-order crossing diagrams.

In the present paper we revisit the problem of the electronic 
structure of the Ce pnictides, and the localization-delocalization phenomena 
of the $4f$ shell, in light of recent developments of treating electron 
correlations directly in electronic structure calculations. 
In particular, we investigate how well dynamical mean field theory reproduces the lattice 
parameters and spectral properties of these materials,
and we have used the implementation reported 
in Refs.~\onlinecite{GrechnFlex,igorMn,patrikHI,oscarDMFTloop} for
these purposes, with the HI and SPTF solvers.

\section{Computational details}

\begin{table}
\caption{Muffin-Tin radii (in atomic units) of the cerium 
and pnictogen atoms used in the present calculations.}
\label{TblRmt}
\begin{ruledtabular}
\begin{tabular}{l|ccccc}
\phantom{x} & CeN & CeP & CeAs & CeSb & CeBi \\
\hline
$R_{MT}$(Ce) & 2.45 & 2.75 & 2.83 & 2.90 & 2.89\\
$R_{MT}$(X) & 2.10 & 2.40 & 2.50 & 2.90 & 2.91\\
\end{tabular}
\end{ruledtabular}
\end{table}

\begin{table}
\caption{Experimental binding energies ($\Delta$E) and
 DC terms for the $4f$ states of cerium in the Hubbard~I approximation.}
 \label{dcTbl}
\begin{ruledtabular}
\begin{tabular}{l|cccc}
\phantom{x} & CeP & CeAs & CeSb & CeBi \\
\hline
$\Delta E$ (eV)  & 2.95 & 3.05 & 3.0 & 3.2 \\
PW91, DC (Ry)  & 0.335 & 0.308 & 0.262 & 0.247 \\
PBE96, DC (Ry) & 0.344 & 0.316 & 0.270 & 0.255 \\
AM05, DC (Ry)  & 0.356 & 0.328 & 0.283 & 0.269 \\
\end{tabular}
\end{ruledtabular}
\end{table}

\begin{table}
\caption{Calculated and experimental lattice parameters of cerium pnictides in atomic units.}
\label{TblLatticeParam}
\begin{ruledtabular}
\begin{tabular}{l|ccccc}
\phantom{x} & CeN & CeP & CeAs & CeSb & CeBi \\
\hline
PW91, $4f$-itinerant & 9.38 & 10.73 & 11.01 & 11.72 & 11.96\\
PBE96, $4f$-itinerant & 9.55 & 10.94 & 11.26 & 12.00 & 12.27\\
AM05, $4f$-itinerant & 9.42 & 10.78 & 11.09 & 11.81 & 12.07\\
\hline
PW91, $4f$-core & 9.79 & 11.08 & 11.31 & 11.92 & 12.10\\
PBE96, $4f$-core & 9.98 & 11.32 & 11.59 & 12.24 & 12.57\\
AM05, $4f$-core & 9.86 & 11.17 & 11.41 & 12.03 & 12.20\\
\hline
PW91+DMFT & 9.38  & 11.09 & 11.32 & 11.95 & 12.13\\
PBE96+DMFT & 9.56 & 11.30 & 11.55 & 12.22 & 12.44\\
AM05+DMFT & 9.43 & 11.19 & 11.43 & 12.06 & 12.25\\
\hline
Experiment & 9.49 & 11.20 & 11.49 & 12.13 & 12.26\\
\end{tabular}
\end{ruledtabular}
\end{table}

The calculations presented in this
work are based on the density functional 
theory (DFT)~\cite{LDAhK,LDAkS}
combined with the dynamical mean-field 
theory~\cite{KatsLiht1,LdaDmft2,
LdaDmft3Georges, LdaDmft4Savrasov, LdaDmft5Held} method.
The DFT calculations were done in the local-density approximation
with exchange-correlation functional PW91~\cite{PW92ref},
and in the generalized gradient approximation
using the PBE96~\cite{PBE96ref} parametrization, 
as well as with the AM05 functional~\cite{AM05ref},
within the full-potential linearised  muffin-tin 
orbitals (FP-LMTO) approach~\cite{wills,LMTO_basic, FPLMTOmeth}. 
The LDA/GGA+DMFT calculations were done by means of the implementation presented in Refs.~\onlinecite{GrechnFlex,igorMn,patrikHI,oscarDMFTloop}, with the total energies evaluated
following Ref.~\onlinecite{igorMn}. To solve the Anderson impurity model arising 
in the DMFT part, the SPTF solver~\cite{flexeqsKatsLicht,GrechnFlex,leonid_sptf} was used
for CeN, whereas the Hubbard~I 
solver~\cite{KatsLiht1,patrikHI,HubbardI}
was used for CeP, CeAs, CeSb, and CeBi.

The LDA/GGA+DMFT scheme was applied to the $4f$ state of cerium
using two types of the correlated orbitals~\cite{GrechnFlex}:
orthogonalized LMTO subspace~(ORT) and muffin-tin only correlated subspace~(MT). 
All these ORT and MT orbitals are localized,
and form orthonormal sets.
The ORT set consists of functions obtained 
from the FP-LMTO method with the 
LDA/GGA Hamiltonian for the correlated states.
The MT correlated subspace is defined 
only into the muffin-tin region, and
outside of the muffin-tin spheres the MT orbitals are taken 
equal to zero by definition.
Properties of these subspaces are analyzed 
in the works published in Refs.~\onlinecite{Haule,GrechnFlex}.
In the present calculations the MT set was used for CeN, and 
the ORT set for all other Ce pnictides.

The LDA/GGA calculations included spin-orbit coupling in the scalar-relativistic approximation,
and only within the muffin-tin spheres. The radii of the muffin-tin spheres
for the different compounds were kept constant while changing the lattice
parameters, and their values are reported in Table~\ref{TblRmt}. For the Brillouin 
zone integration the tetrahedron method was used with
a $16 \times 16 \times 16$ $\mathbf{k}$-mesh.
It corresponds to $85$ $\mathbf{k}$-points in the irreducible
wedge of the Brillouin zone with the total number of $\mathbf{k}$-points 
in the Brillouin zone equal to $2048$.
For the charge density and potential angular decomposition inside
the muffin-tin spheres the value of maximum angular momentum 
was taken equal to $l_{\max}=8$.
Three kinetic energy tails were used, corresponding to 
$\{-0.3, -2.3, -1.6\}$ Ry.
Finally all calculations were non-spin polarized, except CeN which was converged to
a ferromagnetic solution. Magnetic properties were in fact considered at the DMFT level,
and the SPTF solver works better when starting from a magnetic solution~\cite{IgorMn2}.

Calculations without DMFT in the 
pure PW91/PBE96/AM05 approximations were done in two ways:
the cerium $4f$ electrons were treated either as core or as
itinerant states. In the latter case two kinetic energy tails
were considered for the basis.

For the LDA/GGA+DMFT calculations it is necessary to choose an adequate
value for the Hubbard and (Hund) exchange parameters U and J. 
For all Ce pnictides except CeN we have used U=7 eV and J=0.95 eV,
which is compatible with previous calculations~\cite{dmft1CePn}
considering that our method is also based on a fully rotationally invariant
U-matrix~\cite{GrechnFlex}. Conversely the $4f$ states in CeN have
basically an itinerant character, as discussed more in detail in the results
section. This implies that the screening of the Coulomb interaction is
much more effective and the value of U reduced in comparison to the other
Ce pnictides. Following these guidelines and having in mind that the SPTF 
solver tends to overestimate correlation effects~\cite{igorMn,Fe_PRL}
we have used U=2 eV, which is consistent with its applicability
range. Higher values of U have also been studied but these
results have not been reported here as they gave the same
physical picture as U=2 eV. The value of the Hund's exchange is less dependent
on the nature of the screening in a material, and therefore we have used
J=0.95 eV for CeN as well.


To calculate the energy versus volume curves in the Hubbard~I approximation,
it is important to correctly treat 
the double counting term $H_{DC}$ included in the local problem~\cite{igorMn}.
The $H_{DC}$ term is used to remove the wrong contributions
of local Coulomb interaction already contained in the DFT Hamiltonian
projected on the correlated orbitals ($4f$ states) and used in the DMFT part.
The double counting term was first calculated from the HI solver at
experimental lattice constants, and then kept fixed for all the other values.
In order to fix the main occupied peak at the correct position
the photoemission data available from experiments~\cite{xps1, Xray3eV0p3eV, xps3, xps4}
were used. The experimental binding energies and double counting term values are presented
in Table~\ref{dcTbl} for the PW91/PBE96/AM05 + HI approximations.

CeN did not present these problems since in the SPTF solver the double counting 
correction can be directly related to a fully hybridizing self-energy by simply
removing the orbitally averaged $\Sigma (0)$, separately for each spin channel.
When evaluating the local interacting Green's function with the SPTF solver was
important to consider a fully renormalized propagator to obtain
a conserving approximation in Baym-Kadanoff sense~\cite{flex2,flex5,flex6}.

\section{Results}

\begin{figure}[t]
\includegraphics[width=6.5 cm]{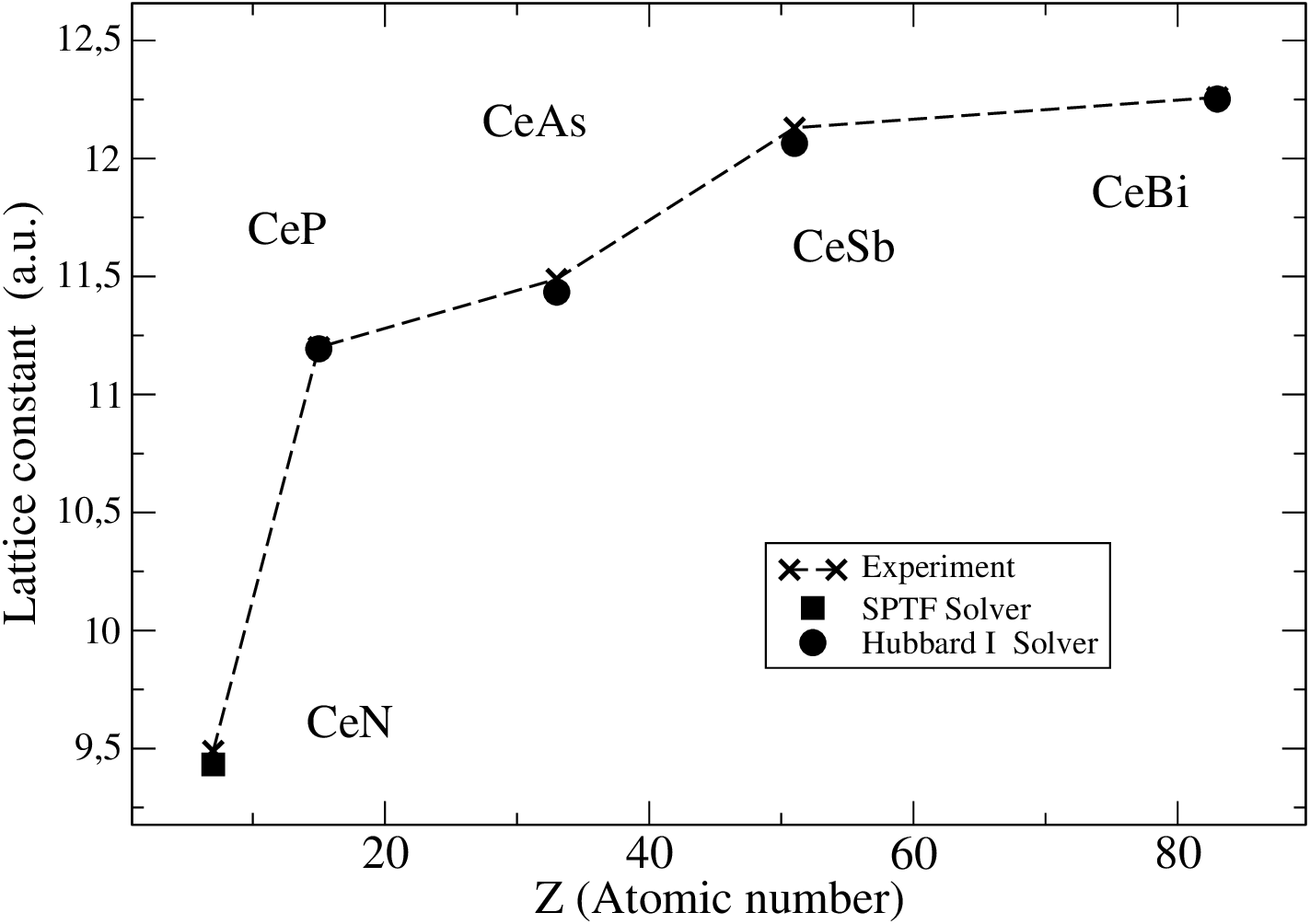}
\caption{Experimental and calculated lattice constants of Ce pnictides. 
The theoretical values for CeN are obtained with the SPTF solver and the AM05 DFT functional.
The Hubbard~I solver together with the AM05 DFT functional was used for all other pnictides.}
\label{Fig1}
\end{figure}

The calculated lattice parameters are presented in
Table~\ref{TblLatticeParam}, where we also show available experimental 
data~\cite{latcons1, Didchenko1963, Svane1998exp,
CeNparamPrccure, BrooksLMTOASA85, latcons2, latcons3, latconsAll}.
A selection of these results are plotted in Fig.~\ref{Fig1}. 
In this figure we show experimental values of lattice constants
and values obtained with 
AM05 DFT functional combined with the SPTF solver for CeN
and with the Hubbard~I solver for all other pnictides. An overall good 
agreement between experimental data and calculations is observed, 
both concerning the trend as well as the absolute values.

\begin{figure}[b]
\includegraphics[width=6.5 cm]{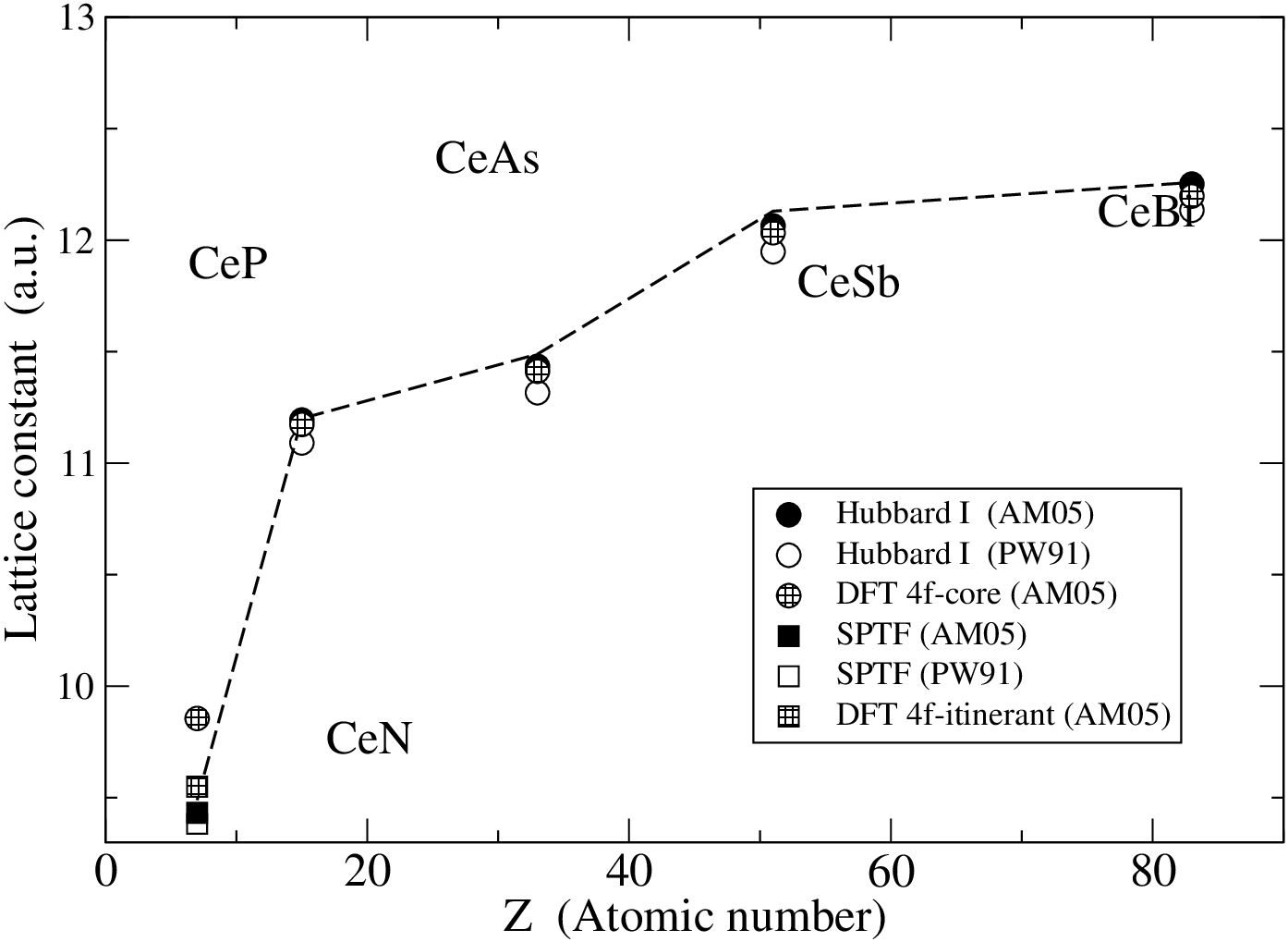}
\caption{Calculated lattice constants of Ce pnictides. The theoretical 
values for CeN are obtained with the SPTF solver and two DFT 
parametrizations, AM05 and PW91. The Hubbard~I solver with both the AM05 and 
PW91 DFT functionals was used for all other pnictides. These data are compared to
calculations using localized $4f$ (core) states with the AM05 and PW91 functional.
For CeN the plain AM05 DFT results with $4f$ valence electrons are also reported,
emphasizing how such a description is much better than the assumption of localized core
states for this material.}
\label{Fig2}
\end{figure}

In Fig.~\ref{Fig2} we present the
calculated lattice constants from different approximations.
Results are given for CeN using the SPTF solver with AM05 and PW91
DFT functionals as well as
for pure AM05 DFT calculations with the
$4f$ electron of Ce atom considered as itinerant.
For all other pnictides, results are given
for the Hubbard~I solver with AM05 and PW91 DFT functionals.
The choice of showing in Fig.~\ref{Fig2} parametrizations 
from the AM05 and PW91 functional was made since
these two approximations give the best and, respectively, 
the worst agreement with observed lattice parameters.
As Fig.~\ref{Fig2} shows, the difference between the two
functionals, when combined with DMFT, is very small and both parametrizations
 reproduce observations with good accuracy.
In Fig.~\ref{Fig2} we also show results 
from more time efficient calculations,
treating the $4f$ states as localized core states and using the AM05 DFT functional.
Moreover for CeN a calculation with the same functional and $4f$ states treated
as itinerant is showed.

\begin{figure}[t]
\includegraphics[width=6.1 cm]{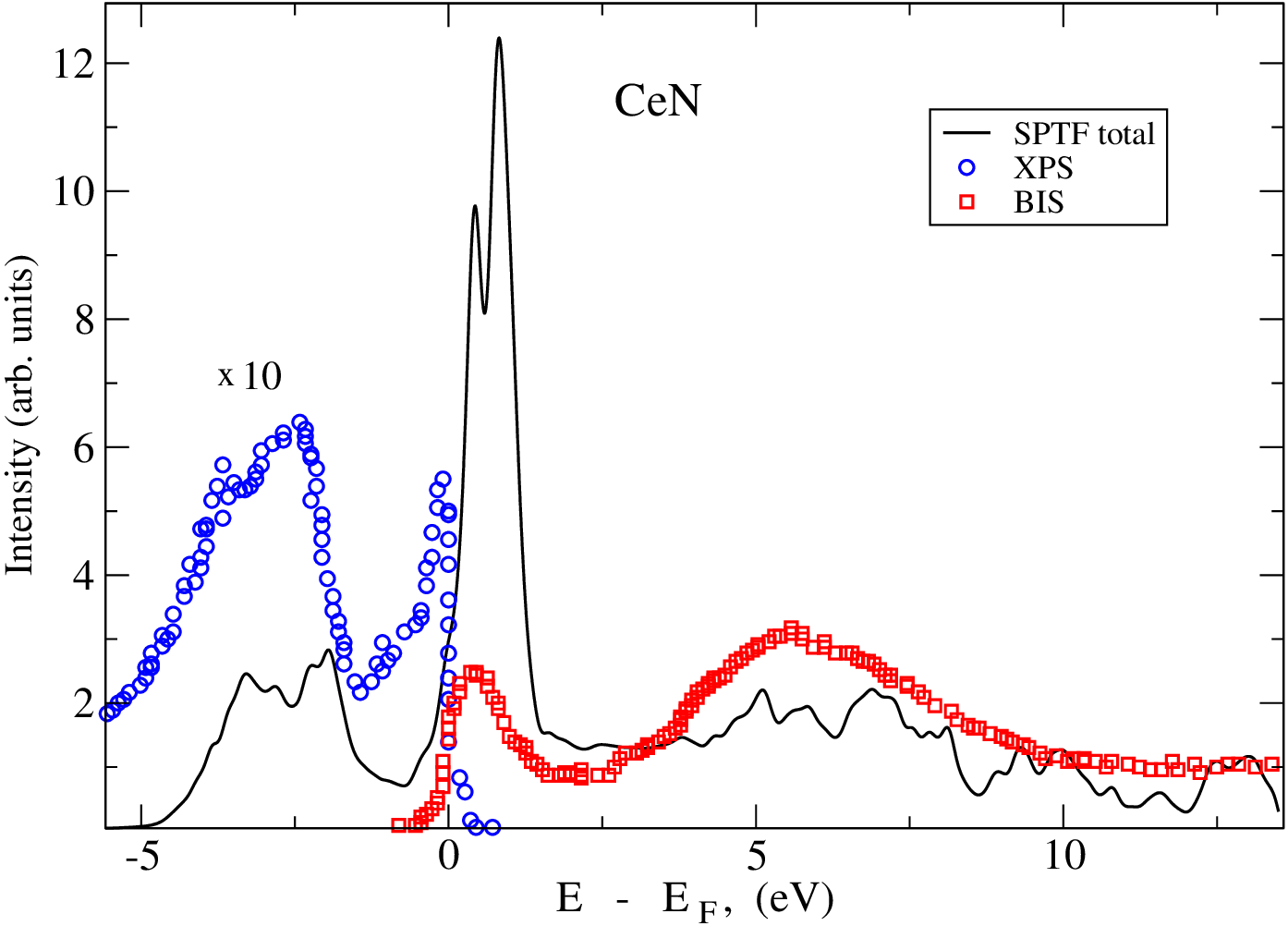}
\caption{(Color online) Experimental and calculated spectrum of CeN.
The blue points label the XPS data and the red points mark the BIS spectrum from Ref.~\onlinecite{cenxpsbis},
where the relative intensities of the XPS and BIS spectra have 
been tentatively adjusted. The factor ($\times 10$) comes directly from Ref.~\onlinecite{cenxpsbis}, and
we have kept it to make a better comparison between the experimental XPS data and the theoretical curve.
The solid line denotes the total spectral function using the SPTF solver for the Ce $4f$ states.}
\label{Fig3}
\end{figure}

The DFT calculations with $4f$ electrons treated
as core states for CeP, CeAs, CeSb and CeBi,
give a good agreement with the experimental data.
On this level of approximation the 
$4f$ states do not contribute to the 
chemical bonding. The Hubbard~I calculations also treat
the $4f$ states as essentially non-bonding and atomic like,
albeit a small degree of hybridization is present 
in such kind of calculations.
All data in Fig.~\ref{Fig1} and Fig.~\ref{Fig2} point to that
the $4f$ states do not contribute to the chemical bonding in the heavier Ce pnictides.

In contrast, the CeN compound is a system which seems 
best described having delocalized $4f$ states as it follows from
the comparison of our calculated values and the experimental data.
The DMFT calculation with the SPTF solver 
and the pure AM05 DFT calculation with itinerant $4f$ states
reproduce the observed lattice constant with good accuracy. Conversely
if one tries to assume a localized $4f$ electron in the form of a core
state the results become more in line with the other Ce pnictides, but
the agreement with experiments worsen. These facts point to that
the $4f$ states have an important role in the chemical bonding, which
results in a significantly smaller lattice constant of CeN in comparison
to the heavier Ce pnictides (Fig.\ref{Fig1}). The delocalized nature
of the $4f$ electron makes it not suitable for the Hubbard I approximation
or any other approach in the strong coupling limit, unless one wants
to apply semi-empirical workarounds. For example, in a previous LDA+DMFT
study~\cite{dmft1CePn} based on the NCA solver, the $4f$ hybridization 
function in CeN had to be rescaled of a factor 2 to avoid an unphysical
occupation number of the $4f$ shell.

To give further evidence to this fact we observe that the
calculated spectral properties of CeN are
in rather good agreement with experimental data, when considering
the $4f$ states as essentially itinerant with some degree of correlation 
resulting from the SPTF solver.
This comparison is presented in Fig.~\ref{Fig3}, where both occupied
and unoccupied states from the calculations are compared to 
experimental XPS and BIS data. Note that the calculation of the occupied 
states reproduce the spectral feature between $3$ and $5$~eV binding energy, 
stemming mostly from N $p$-states, whereas the feature at the Fermi level is 
dominated by the Ce $4f$ states, which form a band-like feature pinned 
at the Fermi level. This results in that the unoccupied states have a $4f$ feature at 
the Fermi level, which is different from the situation for the heavier Ce pnictides, 
as discussed below. The unoccupied states of CeN also have a broader 
feature at $5$ - $6$~eV energy, which is dominated by Ce $5d$ states 
which hybridize with ligand $p$~states.

\begin{figure}[t]
\includegraphics[width=6.1 cm]{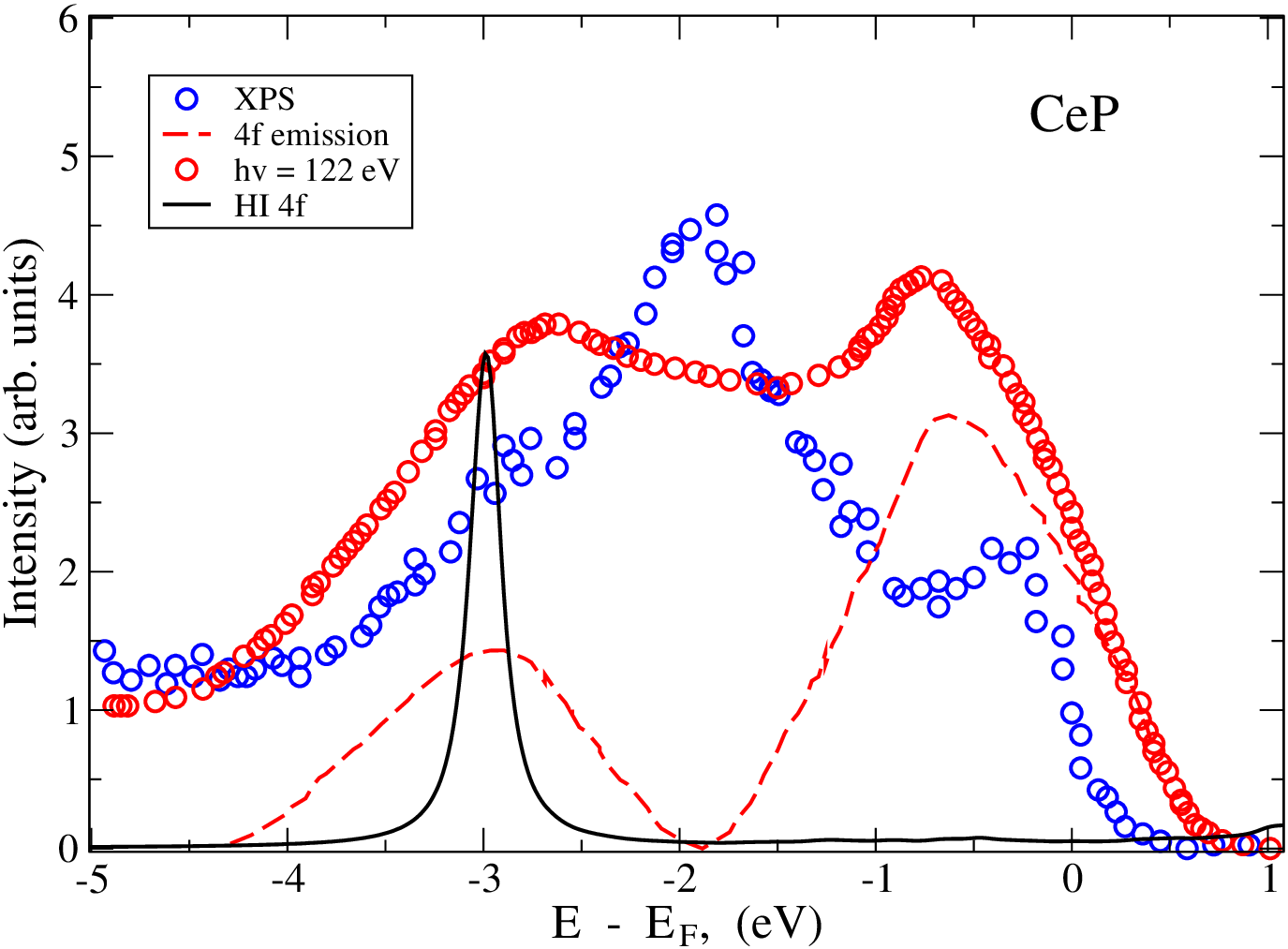}
\caption{(Color online) Experimental and calculated 
spectral properties of the $4f$ states for CeP.
The XPS data from Ref.~\onlinecite{xps1} are marked as blue circles.
The red circles label the resonant spectrum of CeP, while
the red dashed line labels the extracted photoemission spectrum
of $4f$ states from Ref.~\onlinecite{xps2}. 
The black solid line is the calculated spectral function of the Ce $4f$ states
with the Hubbard~I solver.}
\label{Fig4}
\end{figure}
\begin{figure}[b]
\includegraphics[width=6.1 cm]{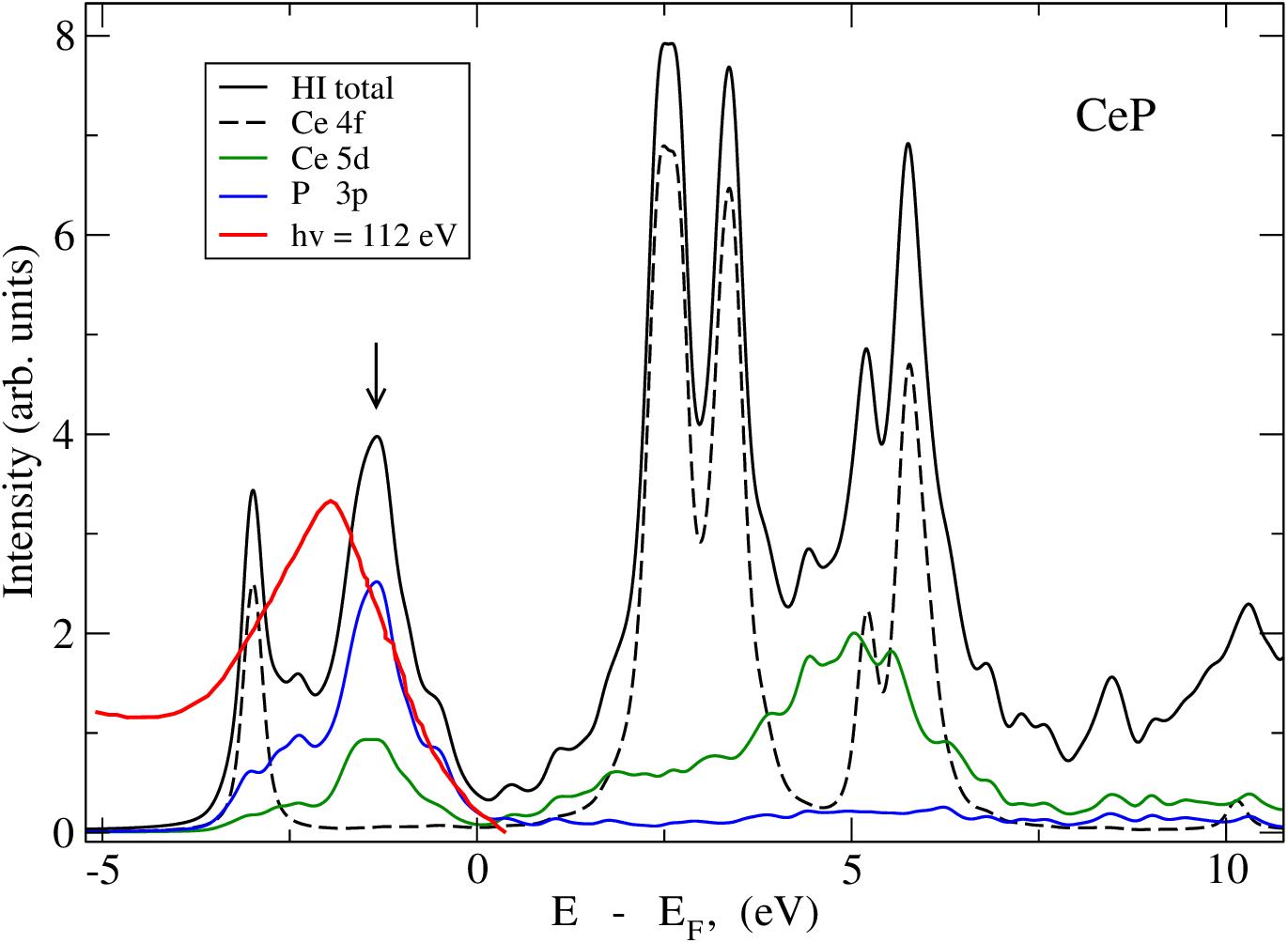}
\caption{(Color online) Total spectral function calculated with the Hubbard~I solver (black solid line), 
compared with experimental data from Ref.~\onlinecite{xps2}.
We also present the projections for the Ce $5d$ and $4f$ states, and the P $3p$ states (dashed lines).
The vertical arrow labels the $pd$ peak which is close to
the experimental non resonant photoemission data taken at $h\nu=112$~eV
(red solid line).}
\label{Fig5}
\end{figure}

In Fig.~\ref{Fig4} we present the CeP $4f$ spectral function, which sometimes is loosely referred to
as the correlated density of states (DOS) to relate it to the band-structure theory. In this figure we
make a comparison to the experimental data from Refs.~\onlinecite{xps1,xps2}.
As one can see, the Hubbard~I approximation
describes the deeply localized $4f$ feature in a proper way, although 
due to life time broadening and finite spectrometer resolution the experimental 
feature is much wider compared to theory. Note that the Hubbard~I approximation 
is not able to capture the well screened feature at the Fermi level, 
and indeed the experimental feature can not be reproduced with the present calculations.

The experimental results for the non resonant frequency of $4f$ Ce states
are compared with the total and partial theoretical spectral functions in Fig.~\ref{Fig5}.
For this photon energy the main contribution to the observed peak
is from cerium $5d$ states and phosphorus $3p$ states~\cite{xps2}. 
The experimental peak-width and position are reproduced by 
the Ce $5d$ and P $3p$ partial spectral functions, as Fig.~\ref{Fig5} shows.

\begin{figure}[t]
\includegraphics[width=6.0 cm]{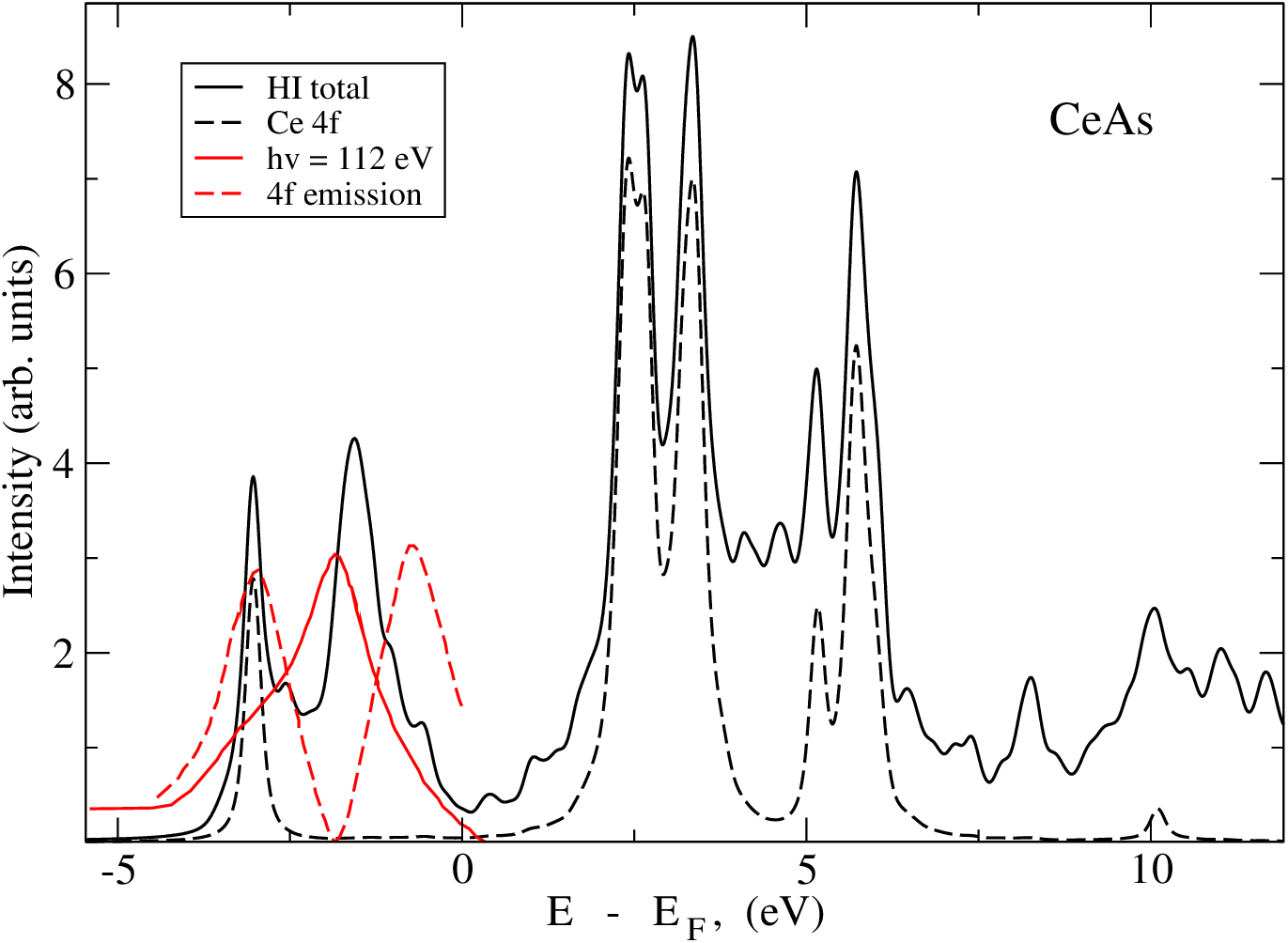}
\caption{(Color online) Experimental and calculated 
spectral properties using Hubbard~I approximation of CeAs.
The total spectral function is labelled by the black solid line, the Ce
$4f$ projection is marked by the black dashed line, while the
red dashed line marks the experimental
resonant $4f$ emission from Ref.~\onlinecite{xps2}.
The red solid line represents the experimental data from nonresonant emission 
(formed by As~$4p$ and Ce~$5d$ states).}
\label{Fig6}
\end{figure}

\begin{figure}[t]
\includegraphics[width=6.0 cm]{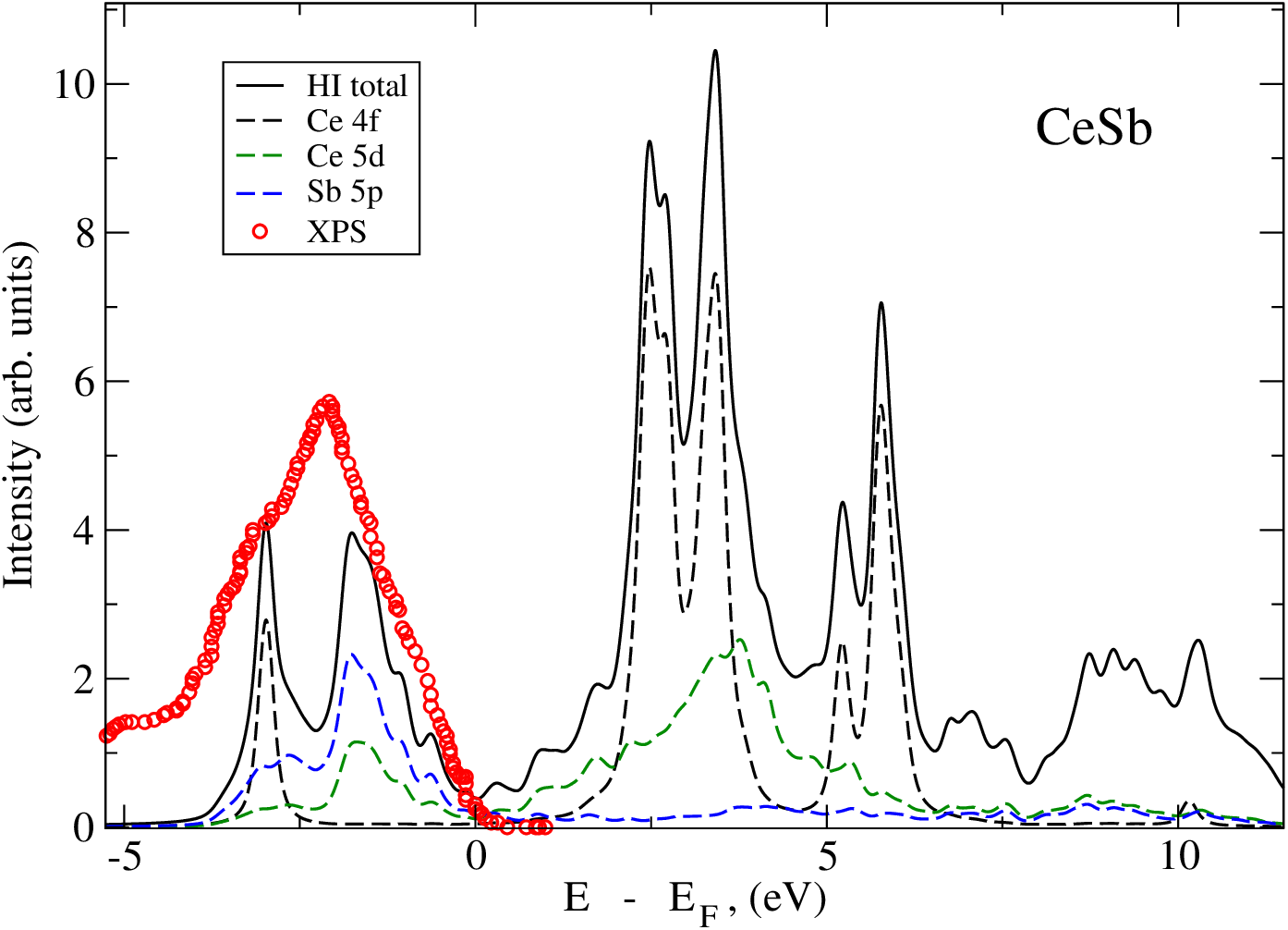}
\caption{(Color online) Total spectral function of CeSb calculated 
with the Hubbard~I solver (black solid line),
compared with the experimental data from Ref.~\onlinecite{xps1}.
We also present projections for the Ce~$4f$ and $5d$ states, and the Sb $5p$~states (dashed lines).
The $pd$-peak is close to the experimental data from nonresonant 
photoemission (red circles).}
\label{Fig7}
\end{figure}

\begin{figure}[b]
\includegraphics[width=6.0 cm]{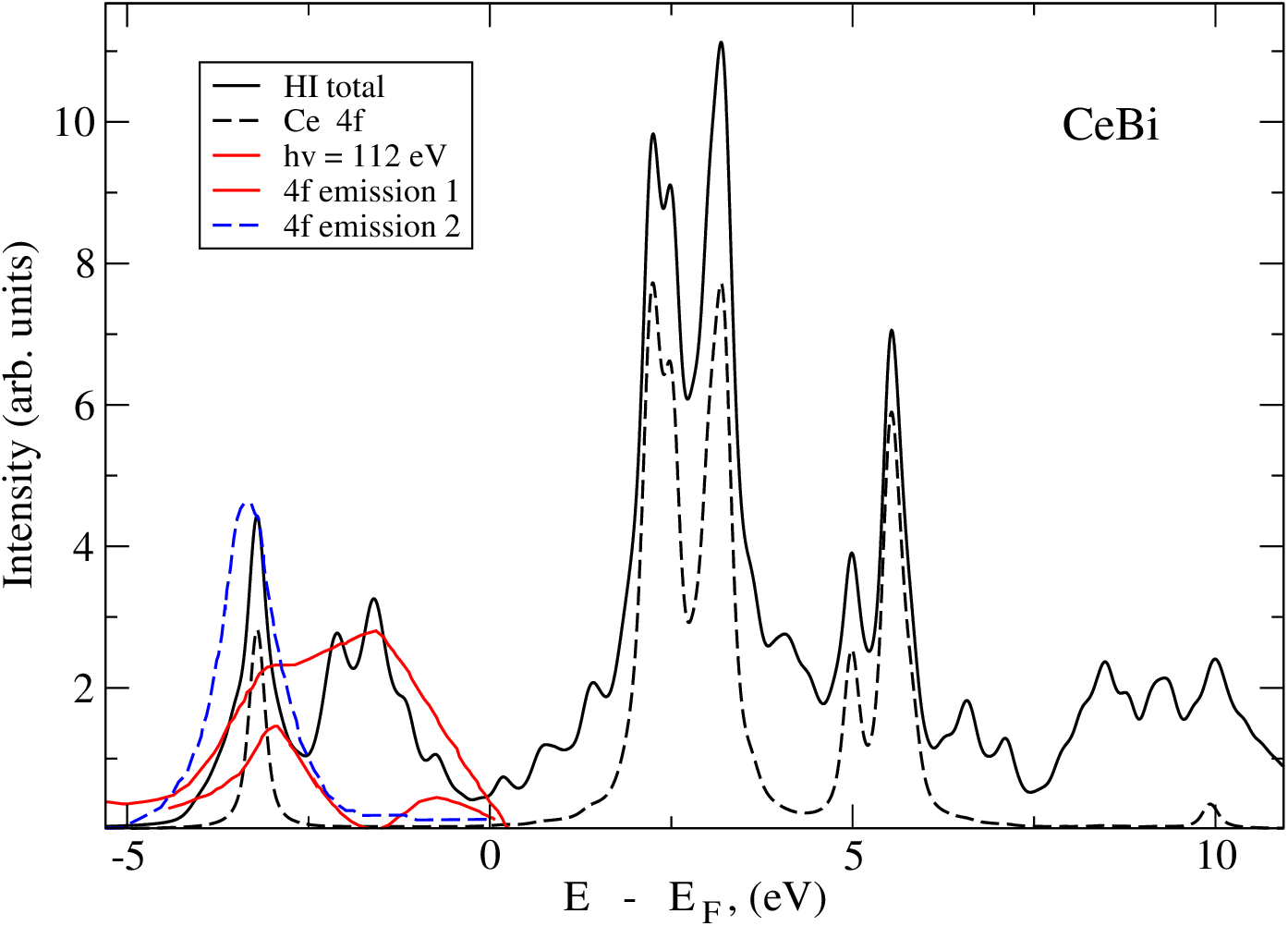}
\caption{(Color online) Experimental and calculated spectral properties of CeBi using the Hubbard~I approximation.
The total spectral function is labelled by the black solid line, while the Ce $4f$ contribution is marked by
the black dashed line. The red dashed line and the red solid line respectively mark the experimental
resonant $4f$ emission from Ref.~\onlinecite{xps2} and the nonresonant emission, formed by the Bi $6p$ and Ce $5d$ states.
The blue dashed line labels the resonant $4f$ emission from Ref.~\onlinecite{xps3}.}
\label{Fig8}
\end{figure}

A similar picture occurs for the other heavy
pnictides, for which we in Figs.~\ref{Fig6},~\ref{Fig7},~\ref{Fig8} compare 
calculated spectral features with experimental data taken on and off resonance. 
As discussed, the $4f$ feature close to the
Fermi energy (the well screened peak) is not reproduced in our
Hubbard~I calculation. However, 
going from CeP to CeBi the experimental
$4f$ peak at the Fermi level decreases in intensity
and becomes almost negligible in the CeBi case.
For this pnictide the Hubbard~I approximation hence
gives the best description of the spectral features.

In Figs.~\ref{Fig5},~\ref{Fig6},~\ref{Fig7}, and~\ref{Fig8} we also show 
the unoccupied states. It is clear that the $4f$ dominated spectral 
features are due to an $f^1 \to f^2$ transition, 
in which multiplet effects are clearly visible. Since we have not found 
experimental BIS data for the heavy Ce pnictides our theory must be viewed
as a prediction. We note that a calculation based on the LDA or LDA+U approximation
would not result in spectral features displaying multiplet effects, and an experimental 
verification or rebuttal of the BIS data shown 
in Figs.~\ref{Fig5},~\ref{Fig6},~\ref{Fig7},~\ref{Fig8} would give valuable 
insight into which approximation is best for these systems. 
On general grounds however, one would expect for these materials that
the Hubbard~I approximation is better than LDA or LDA+U.


A similarity can be observed between the heavier Ce pnictides and $\gamma$-Ce.
In the photoelectron spectrum of Ce both the $\gamma$-phase 
and the $\alpha$-phase display the well screened $4f$~feature and the poor 
screened $4f$~feature, which for $\gamma$-Ce is located at about $2.0$ eV binding energy \cite{xps3,dmftpeak}. 
$\gamma$-Ce is a localized $4f$ system, 
and therefore a Hubbard~I calculation for $\gamma$-Ce 
should reproduce the lattice constant with equal accuracy 
as it does for the Ce pnictides. Our calculations for $\gamma$-Ce actually 
confirm this expectation and we obtain an equilibrium lattice constant of
$9.56$~a.u., $9.95$~a.u, $9.78$~a.u., respectively for PW91+HI, PBE96+HI, and 
AM05+HI. These values should be compared to the
experimental lattice parameter which is equal to $9.75$~a.u.~\cite{ceconstant},
showing again that the AM05+HI approach results in the best agreement with experiment.
For $\gamma$-Ce we have also calculated the bulk modulus to investigate if the lack of
the well-screened $4f$~feature in the spectrum leads to a deficiency in more sensitive ground
state properties than the lattice constant. We obtain a bulk modulus of $37$~GPa, $31$~GPa, and
$27$~GPa, respectively for PW91+HI, PBE96+HI, and AM05+HI. These values are in line with
the GGA+U result of $34$~GPa~\cite{amadon_gamma_ce}, and overestimate the experimental
bulk modulus, which is equal to $19$~GPa\cite{amadon_gamma_ce}. A better theory based
on a more sophisticated approach would probably be able to correct this disagreement 
with the experiments.

\section{Conclusions}
We conclude that the LDA/GGA+DMFT approach
used in the present work with the SPTF and Hubbard~I
solvers gives very good values of lattice constants 
and acceptable spectral properties of Ce pnictides.
The Hubbard~I solver is not able to capture the experimental 
Kondo-like feature at the Fermi level observed for CeP, CeAs and
CeSb, while for CeBi this peak is almost absent, and only a small shoulder
is visible. This deficiency is inner to the method used, and
may potentially  be remedied with more sophisticated techniques
such as quantum Monte-Carlo. However, in this case, technical problems of 
different nature may be introduced and related to the statistical noise of 
the Green's function. For reasonable computational resources the occupation
numbers and the total energies would be less stable, and the spectral function
would suffer of the uncertainty related to the maximum entropy method
for analytical continuation. Finally it would also be problematic to include 
the spin-orbit coupling without arbitrarily discarding parts of the hybridization function.

It is interesting, nevertheless, that despite the inability of the Hubbard~I approximation
(or the SPTF solver) to capture the well-screened Kondo-like peak
observed for some Ce pnictides the lattice constants are reproduced with a very
good accuracy. This implies that the mechanism behind this feature
does not contribute much to the chemical bonding. More sensitive properties as, e.g., the 
bulk modulus would probably show a signature of this discrepancy between theory
and experiment, as verified in this paper for $\gamma$-Ce. It would be of a great
interest to approach this problem with more refined methods, and this will
be the aim of our future research.

Finally we also note that the lattice constants for the heavier Ce pnictides, 
in which the $4f$ states are essentially localized and non-bonding, 
can be well reproduced from a theory where these states are treated
as core-states where they are not allowed to hop from lattice site to lattice site, 
and where hybridization to all other states is absent. Hence, a good alternative
to treat the $4f$ states of localized $4f$ electron systems is to simply consider
them as core like, which is an approximation that offers great computational 
speed with decent accuracy.

\section*{ACKNOWLEDGMENTS}
We are very grateful to N.~M{\aa}rtensson
for the valuable discussions of the  
experimental data from the Ref.~\onlinecite{xps2}
and to M.I.~Katsnelson and K.~Held for discussions regarding the
theoretical aspects of this  work. 
The computations were performed on resources  provided
by the Swedish National Infrastructure for Computing (SNIC)
at Uppsala Multidisciplinary Center
for Advanced Computational Science (UPPMAX)
and on resources provided by SNIC
at the National Supercomputer Center (NSC)
under the Project \textit{matter2}.
We acknowledge support from the Swedish 
Research Council. O.E. acknowledges 
support from the KAW foundation as 
well as the ERC (ASD - project 247062).


\begin{thebibliography}{71}
\bibitem{johansson} B. Johansson, Phil. Mag. \textbf{30}, 469 (1974).
\bibitem{allen} J.M. Allen and R.M. Martin, Phys. Rev. Lett. \textbf{49}, 1106 (1982). 
\bibitem{koskimaki} D.C. Koskimaki, K.A. Gschneidner, Phys. Rev. B \textbf{11}, 4463 (1975). 
\bibitem{tsymbal}Ch.-G. Duan, R.F. Sabirianov, W.N. Mei,
P.A. Dowben, S.S. Jaswal, and E.Y. Tsymbal,
J. Phys.: Condens. Matter \textbf{19}, 315220 (2007).
\bibitem{Didchenko1963}
R. Didchenko and F.P. Gortsema, J. Phys. Chem. Solids \textbf{24}, 863 (1963).
\bibitem{patthey1}
F. Patthey, S. Cattarinussi, W.-D. Schneider, Y. Baer, and B. Delley,
Europhys. Lett. \textbf{2}, 883 (1986).
\bibitem{patthey2}
F. Patthey, J.-M. Imer, W.-D. Schneider, H. Beck, and Y. Baer,
Phys. Rev. B {\bf 42}, 8864 (1990). 
\bibitem{delin} A. Delin, P. M. Oppeneer, M.S.S. Brooks, 
T. Kraft, J.M. Wills, B. Johansson, and O. Eriksson, 
Phys. Rev. B \textbf{55}, R10173 (1997).
\bibitem{schoenes} J. Schoenes, in Handbook on the Physics and Chemistry of the Actinides, edited by A.J. Freeman and G.H. Lander~North-
Holland, Amsterdam, 1984, Vol. 1, Chap. 5; 
J. Schoenes, in Moment Formation in Solids. Proceedings of the Nato Advanced
Study Institute, edited by W.J.L.~Buyers, Plenum, New York, 1984, p. 237.
\bibitem{birgenau} R.J. Birgeneau, E. Bucher, J.P.~Maita, L.~Passell, K.C.~Turberfield,
Phys. Rev. B \textbf{8}, 5345 (1973).
\bibitem{cooper1} B.R. Cooper, R. Siemmann, D. Yang, P. Thayamballi, 
and A. Banerjea, in Handbook on the Physics and Chemistry of the
Actinides, edited by A.J. Freeman and G.H. Lander 
(North-Holland, Amsterdam, 1985), Vol. 2, Chap. 6, p. 435.
\bibitem{cooper2} B.R. Cooper, J. Less Common Met. \textbf{133}, 31 (1987).
\bibitem{rossa}J. Rossat-Mignod, D. Delacote, J.M. Effantic, C. Vettier, O. Vogt,
Physica B, \textbf{120}, 163 (1983).
\bibitem{vachter}A. Schlegel, E. Kaldis, P. Wachter, and Ch. Zurcher,
Phys. Lett. A \textbf{66}, 125 (1978).
\bibitem{xps12}
Y. Baer and Ch. Z\"urcher,
Phys. Rev. Lett. \textbf{39}, 956 (1977).
\bibitem{xps1}
Y. Baer, R. Hauger, Ch. Z\"urcher,
M. Campagna, G.K.~Wertheim,
Phys. Rev. B \textbf{18}, 4433 (1978).
\bibitem{xps2} A. Franciosi, J.H. Weaver, N. M{\aa}rtensson, M. Croft,
Phys. Rev. B \textbf{24}, 3651 (1981).
\bibitem{xps3}
J.W. Allen, S.-J. Oh, I. Lindau, J.M. Lawrence, L.I. Johansson,
and S.B. Hagstr\"om,
Phys. Rev. Lett. \textbf{46}, 1100 (1981).
\bibitem{xps4}
H. Kumigashira, H.-D. Kim, A. Ashihara, A. Chainani,
T. Yokoya, T. Takahashi, A. Uesawa, and T. Suzuki, 
Phys. Rev.~B~\textbf{56}, 13654 (1997).
\bibitem{xps5}
M. Croft, A. Franciosi, J.H. Weaver, A. Jayaraman,
Phys. Rev. B \textbf{24}, 544 (1981) 
\bibitem{xps6}
W. Gudat, M. Campagna, R. Rosei, J.H. Weaver, W. Eberhardt, F. Hulliger, and E. Kaldis,
J. Appl. Phys. \textbf{52}, 2123 (1981).
\bibitem{GSmodel}
O. Gunnarsson and K. Schonhammer, Phys. Rev. B
\textbf{28}, 4315 (1983).
\bibitem{twofpeak1}
J.C. Fuggle, M. Campagna, Z. Zolnierek, R. L\"asser, and A. Platau,
Phys. Rev. Lett. \textbf{45}, 1597 (1980).
\bibitem{twofpeak2}
J.W. Allen, S.-J. Oh, M.B. Maple, M.S. Torikachvili, 
Phys. Rev. B \textbf{28}, 5347 (1983).
\bibitem{twofpeak3}
N. M{\aa}rtensson, B. Reihl, R.D. Parks,
Solid State Communications \textbf{41}, 573 (1982).
\bibitem{twofpeak4}
M.R. Norman, D.D. Koelling, A.J. Freeman, H.J.F.~Jansen, B.I. Min, T. Oguchi, and Ling Ye,
Phys. Rev. Lett. \textbf{53}, 1673 (1984).
\bibitem{twofpeak5}
S.H. Liu, K.-M. Ho, Phys. Rev. B \textbf{28}, 4220 (1983).
\bibitem{cen2011Olle}
V. Kanchana, G. Vaitheeswaran, Xinxin Zhang, Yanming Ma, A. Svane, and O. Eriksson,
Phys. Rev. B \textbf{84}, 205135 (2011) 
\bibitem{gw10}
T.~Kotani, M.~van~Schilfgaarde, and S.V.~Faleev, 
Phys. Rev.~B~\textbf{76}, 165106 (2007).
\bibitem{wills}J.M. Wills and B.R. Cooper, Phys. Rev. B \textbf{36}, 3809 (1987).
\bibitem{dmft1CePn}
J. Laegsgaard and A. Svane, 
Phys. Rev. B \textbf{58}, 12817 (1998).
\bibitem{dmft2CePn}
O. Sakai and Yu. Shimizu, J. Phys. Soc. Japan \textbf{76}, 044707 (2007).
\bibitem{dmft3CePn}
O. Sakai, Yu. Shimizu, and Ya. Kaneta,
J. Phys. Soc. Japan \textbf{74}, 2517 (2005).
\bibitem{GrechnFlex}
A. Grechnev, I. Di Marco, M.I. Katsnelson, A.I. Lichtenstein,
J. Wills, and O. Eriksson, Phys. Rev. B \textbf{76}, 035107 (2007).
\bibitem{igorMn}I.~Di~Marco, J.~Minar, S.~Chadov, M.I.~Katsnelson,
H.~Ebert, and A.I.~Lichtenstein,
Phys. Rev. B \textbf{79}, 115111 (2009).
\bibitem{patrikHI}P.~Thunstr\"om, I.~Di~Marco, A. Grechnev,
S.~Lebegue, M.I.~Katsnelson, A.~Svane, and O.~Eriksson,
Phys. Rev. B \textbf{79}, 165104 (2009).
\bibitem{oscarDMFTloop} O. Gr{\aa}n\"as, I.~Di~Marco, P.~Thunstr\"om,
L.~Nordstr\"om, O.~Eriksson,  T.~Bj\"orkman, J.M.~Wills,
Computational Materials Science \textbf{55}, 295 (2012).

\bibitem{LDAhK}
P. Hohenberg and W. Kohn, 
Phys. Rev. \textbf{136}, B864 (1964).
\bibitem{LDAkS}
W. Kohn and L. Sham,
Phys. Rev. \textbf{140}, A1133 (1965).
\bibitem{KatsLiht1}
A.I. Lichtenstein and M.I. Katsnelson,
Phys. Rev. B \textbf{57}, 6884 (1998).
\bibitem{LdaDmft2}
V.I. Anisimovy, A.I. Poteryaevy, M.A.~Korotiny, A.O.~Anokhiny, and
G. Kotliar, J.~Phys. Condens. Matter \textbf{9}, 7359 (1997).
\bibitem{LdaDmft3Georges}
A. Georges, G. Kotliar, W. Krauth, and M.J. Rozenberg,
Rev. Mod. Phys. \textbf{68}, 13 (1996).
\bibitem{LdaDmft4Savrasov}
G. Kotliar, S.Y. Savrasov, K. Haule, V.S. Oudovenko,
O.~Parcollet and C.A. Marianetti,
Rev. Mod. Phys. \textbf{78}, 865 (2006).
\bibitem{LdaDmft5Held}
K. Held, Adv. Phys. \textbf{56}, 829 (2007).
\bibitem{PW92ref}
J.P. Perdew and Y. Wang, 
Phys. Rev. B \textbf{45}, 13244 (1992).
\bibitem{PBE96ref}
J.P. Perdew, K. Burke, M. Ernzerhof,
Phys. Rev. Lett. \textbf{77}, 3865 (1996).
\bibitem{AM05ref}
R. Armiento and A.E. Mattsson,
Phys. Rev. B \textbf{72}, 085108 (2005).
\bibitem{LMTO_basic}
O.K. Andersen, Phys. Rev. B \textbf{12}, 3060 (1975).
\bibitem{FPLMTOmeth}
J.M.~Wills, O.~Eriksson, M.~Alouani, and D.L.~Price,
"Full-Potential LMTO Total Energy and Force Calculations"
in \textit{Elecytronic structure and physical properties of solids}
Ed. Hugues Dreysse, Springer Verlag, Berlin (2000) p.148;
J.M.~Wills, M.~Alouani, P.~Andersson, A.~Delin,
O.~Eriksson, O.~Grechnev,
``Full-Potential Electronic structure method''
Springer, Berlin, (2010).
\bibitem{flexeqsKatsLicht}
M.I. Katsnelson and A.I. Lichtenstein,
Eur. Phys. J. B, \textbf{30}, 9 (2002).
\bibitem{leonid_sptf}
L.V. Pourovskii, M.I. Katsnelson, and A.I. Lichtenstein,
Phys. Rev. B \textbf{72}, 115106 (2005).
\bibitem{HubbardI}
J. Hubbard, Proc. R. Soc. London, Ser. A \textbf{276}, 238 (1963);
J. Hubbard, Proc. R. Soc. London, Ser. A \textbf{277}, 237 (1964);
J. Hubbard, Proc. R. Soc. London, Ser. A \textbf{281}, 401 (1964);
J. Hubbard, Proc. R. Soc. London, Ser. A \textbf{285}, 542 (1965).
\bibitem{Haule}
K. Haule, Ch.-H. Yee, and K. Kim,
Phys. Rev. B \textbf{81}, 195107 (2010).
\bibitem{IgorMn2}
I. {Di Marco}, J. Min\'ar, J. Braun, M.I. Katsnelson, A. Grechnev, H. Ebert, A.I. Lichtenstein, and O. Eriksson,
Eur. Phys. J. B \textbf{72}, 473 (2009).
\bibitem{Fe_PRL}
J. S\'anchez-Barriga, J. Fink, V. Boni, I. Di Marco, J. Braun, J. Min\'ar, A. Varykhalov, O. Rader, V. Bellini, F. Manghi, H. Ebert, M.I. Katsnelson, A.I. Lichtenstein, O. Eriksson, W. Eberhardt, and H. A. D\"urr
Phys. Rev. Lett. \textbf{103}, 267203 (2009)
\bibitem{Xray3eV0p3eV}
H. Kumigashira, S.-H. Yang, T. Yokoya, A. Chainani, T. Takahashi, A. Uesawa, and T. Suzuki, 
Phys. Rev. B \textbf{55}, R3355 (1997).
\bibitem{flex2}
P.G. Baym and L. Kadanoff
Phys. Rev. B \textbf{124}, 287 (1961).
\bibitem{flex5}
V. Drchal, V. Janis, J. Kudrnovsky, V.S. Oudovenko, X.~Dai, K. Haule, and G. Kotliar,
J. Phys.: Condens. Matter \textbf{17}, 61 (2005).
\bibitem{flex6}
I. {Di Marco}, P. Thunstr\"om, L. Pourovskii, O. Gr{\aa}n{\"a}s, L. Nordstr\"om, M. I. Katsnelson, and O. Eriksson,
{\itshape{in preparation}}.

\bibitem{ceall}
W. Gudat, R. Rosei, J.H. Weaver, E. Kaldis and F. Hulliger,
Solid State Comm. \textbf{41}, 37 (1982).
\bibitem{latcons1}
J.M. Leger, D. Ravot, J. Rossat-Mignod,
J. Phys. C. Solid State Phys.~\textbf{17}, 4935 (1984).
\bibitem{Svane1998exp} 
A. Svane, Z. Szotek, W.M. Temmerman, J. Lagsgaard, and H. Winter,
J. Phys. Condens. Matter \textbf{10}, 5309 (1998).
\bibitem{CeNparamPrccure}
G.L. Olcese, J. Phys. F: Metal Phys. \textbf{9}, 569 (1979).
\bibitem{BrooksLMTOASA85}
M.S.S. Brooks, J. Magn. Magn. Mat. \textbf{47}, 260 (1985).
\bibitem{latcons2}
F. Hulliger and H.R. Ott, Z. Physik B \textbf{29}, 47 (1978).
\bibitem{latcons3}
R.P.C. Schram, J.G. Boshoven, E.H.P. Cordfunke,
R.J.M. Konings, R.R. van der Laan,
J. Alloys and Comp.~\textbf{252}, 20 (1997).
\bibitem{latconsAll}
Ch.-G. Duan, R.F. Sabirianov, W.N. Mei, P.A. Dowben,
S.S. Jaswal, and E.Y. Tsymbal,
J. Phys.: Condens. Matter~\textbf{19}, 315220 (2007).
\bibitem{cenxpsbis}
E. Wuilloud, B. Delley, W.-D. Schneider, Y. Baer,
J. Magn. Magn. Mat.~\textbf{47}, 197 (1985).
\bibitem{dmftpeak}
D.M. Wieliczka, C.G. Olson, and D.W. Lynch,
Phys. Rev.~B~\textbf{29}, 3028 (1984).
\bibitem{ceconstant}
B.J.~Beaudry and P.E.~Palmer, J.~Less-Common Metals
\textbf{34}, 225 (1974); D.C.~Koskimaki,
K.A.~Gschneider~Jr., and N.T.~Panousis,
J.~Crystal Growth \textbf{22}, 225 (1974).
\bibitem{amadon_gamma_ce}
B. Amadon, F. Jollet, and M. Torrent,
Phys. Rev. B \textbf{77}, 155104 (2008)

\end{thebibliography}
\end{document}